\documentclass[12pt,A4paper]{article}

\usepackage{amsthm,amsmath,natbib}
\RequirePackage[colorlinks,citecolor=blue,urlcolor=blue]{hyperref}
\usepackage{epsfig,multicol,multirow}
\usepackage{geometry}
\def\e{\hbox{E}}
\def\IF{\hbox{IF}}
\def\cov{\hbox{Cov}}
\def\var{\hbox{Var}}
\def\corr{\hbox{Corr}}

\def\se{\hbox{SE}}

\def\min{\hbox{min}}

\renewcommand{\baselinestretch}{1.5}

\begin{document}

\title{When large n is not enough---Distribution-free Interval Estimators for Ratios of Quantiles}

\author{Luke A. Prendergast\thanks{luke.prendergast@latrobe.edu.au} and Robert G. Staudte\thanks{Corresponding author: r.staudte@latrobe.edu.au}\\
Department of Mathematics and Statistics, La Trobe University\\
Melbourne, Victoria, Australia, 3086}
\date{13 September, 2015}
\maketitle

\begin{abstract}
Ratios of sample percentiles or of quantiles based on a single sample are often published for skewed income data to illustrate aspects of income inequality, but distribution-free confidence intervals for such ratios are not available in the literature. Here we derive and compare two large-sample methods for obtaining such intervals. They both require good distribution-free estimates of the quantile density at the quantiles of interest, and such estimates have recently become available. Simulation studies for various sample sizes are carried out for Pareto, lognormal and exponential distributions, as well as fitted
 generalized lambda distributions, to determine the coverage probabilities and widths of the intervals. Robustness of the estimators to contamination or a positive
proportion of zero incomes is examined via influence functions and simulations. The motivating example
 is Australian household income data where ratios of quantiles measure inequality, but of course these  results apply equally to data from other countries.
\end{abstract}


\clearpage
\newpage

\section{Introduction}\label{sec:intro}

Ratios of percentiles from a single population may be of direct interest for many disciplines, but in particular it is very often used as a simple measure of income inequality.  For example, in a recent brief discussion of income inequality measures, \cite{DE07} remarks that decile ratios are simple but effective.
Also, in the inequality literature one often finds  estimated ratios of quantiles plotted against the year in which the samples are taken, to illustrate whether inequality is growing or decreasing over time.
Of course such plots can be misleading, and what is required are inferential methods based on sample ratios.

Recent contributions to inference for ratios of quantiles based on {\em two} independent samples
are found in \cite{bp-2002}, \cite{chwu-2010}. However, to our knowledge there are no published results
based on inference for ratios of quantiles based on a single sample.

The main results presented in this paper are first, showing that large sample distribution-free confidence intervals for ratios of quantiles based on standard theory have reliable coverage for moderate sample sizes.
 Second, even for samples of size 10,000, the standard errors of ratio estimators cannot be ignored; thus one cannot assume that sample ratios are accurate just because the sample size is \lq large\rq .\
Third, showing such procedures are resistant to outliers and to the presence of a small proportion of zero incomes in the population.
 The  same cannot be said for most inequality measures, as shown by \cite{cowell-1996}; although progress in robustifying some of them has been achieved, see \cite{cowell-2003} and references therein.

In Section~\ref{sec:ABS data} we examine income data from the Australian Bureau of Statistics, and
illustrate how our results can provide useful inferential information.
 Then in Section~\ref{sec:DFcis} we find distribution free-standard errors for the ratio of quantiles which require distribution-free estimates of the quantile density at the two quantiles defining the ratio.
   Two interval estimators are described, one based on the studentized log-transformation and the other on a variance stabilization transformation.  Simulation studies in Section~\ref{sec:simulations} show that these intervals rarely have coverage below the nominal level for several distributions that are commonly assumed for income populations, and that the intervals based on variance stabilization  have more accurate coverage and smaller widths for small to moderate sample sizes. Similar good results are obtained for data fitted by the generalized lambda distribution; these are relegated to the Appendix, Section~\ref{sec:GLD}. In Section~\ref{sec:contamination} the effects of contamination by a point mass of zero incomes or infinitesimal contamination are studied via simulations and influence functions. In Section~\ref{sec:twotest} interval estimators for the difference between two independent ratios are found effective.
   The software R script for computing the intervals is found in Section~\ref{Rscript}, and further research is suggested in Section~\ref{sec:summary}.

\section{Australian Bureau of Statistics income data}\label{sec:ABS data}
Measuring household income is a complicated task carried out by governmental departments, including the Australian Bureau of Statistics, whose annual reports are available at \cite{ABSincomedata}.
The gross household income per week is of interest but  households differ so much in size
that the {\em equivalized} disposal income (EWI) is also found, and which the ABS
defines as  \lq ... the amount of disposable cash income that a single person household would require to maintain the same standard of living as the
household in question, regardless of the size or composition of the latter.\rq\

\marginpar{Table 1 here.}

In Table~\ref{table1} we list ratios of percentiles from \cite{ABSincomedata}.
 While details of how the percentiles were calculated are not reported, the sample size of households each year ranges from 9,345 for 2007 and 18,071 for 2009.

\marginpar{Figure 1 here.}

In Figure~\ref{fig1} are histograms of the EWI data for the financial years beginning 2005 and
2011, listed in Table~\ref{table7} of the Appendix, after exclusion of 0 income and income greater than \$2000. Below them are density plots of the \lq reconstructed\rq\  data sets from which we will obtain our quantile estimates and standard errors. Superimposed on the density
plots are gamma densities whose parameters are estimated by the method of moments. We are not
advocating these gamma models for inference regarding quantiles, but rather we generate
random samples from them to assist in assessing interval coverage for such data.

In Table \ref{table2} we report the results for our reconstructed data.  VST and `Stud' refer to 95\% interval estimators that we introduce soon in Section \ref{subsec:DFcis} based on variance stabilization and studentization respectively.  As can be seen the studentized and VST intervals are identical to two decimal places which is due to the large sample sizes.  It should be noted that the widths of the intervals are not so narrow so as to make the intervals redundant, but rather they are themselves informative when reported with the ratio point estimates.  Also shown are results for 10,000 simulation runs from gamma distributed data with parameters set to those used to overlay the densities in Figure \ref{fig1}. They indicate excellent coverage of the intervals when sampling from the fitted gamma distributions with approximately the same results for both methods.

\marginpar{Table 2 here.}

Let $z_\alpha$ denote the $100\alpha$\% quantile from the standard normal distribution.  A test of $\rho _\text{2005}=\rho _\text{2011}$ for the ratio P90/P10 against a significant difference would reject at level 0.05 when $S= |\ln (\hat \rho _\text{2005})-\ln (\hat \rho _\text{2011})| \geq z_{0.975} \times \se $, where
$\se =\se [\ln (\hat \rho _\text{2005})-\ln (\hat \rho _\text{2011})]=\{\se _\text{2005}^2+\se _\text{2011}^2\}^{1/2}$\,,
and $\se _\text{2005}=\se [\ln (\hat \rho _\text{2005})]=0.0105$ and $\se _\text{2022}=\se [\ln (\hat \rho _\text{2011})]=0.0088.$ Now $|S/\se |=2.32 >1.96$, so the P90/P10 ratios differ significantly for the years 2005 and 2011.
Formalities are given in Section~\ref{sec:twotest}.

\section{Distribution-free confidence intervals}\label{sec:DFcis}

\subsection{Distribution-free standard errors for ratios of quantiles}\label{sec:DFses}

Let $F$ be a continuous distribution with positive domain. Define the {\em quantile function} as the inverse $G(p)=F^{-1}(p)=\inf \{x:\ F(x)\geq p\}$, $0<p<1$.  When $F$ is understood, we often write $x_p$
for the $q$th quantile $G(p)$, which is also called the $100p$th percentile. For any choices of $p\neq q$ in $(0,1)$ we are interested in the ratios
\begin{equation}
\rho=\rho(p,q)=\frac{x_p}{x_q}~.\label{rho}
\end{equation}

One can estimate the $p$\,th quantile $x_p=G(p)$ by  $X_{([np]+1)}$,  the $[np]+1$st order statistic of a sample of size $n$ from $F$. However, the \cite{hynd-1996}
 quantile estimator $\hat x_p$, which is a linear combination of two adjacent order statistics, generally has much  less bias and similar variance, so in the sequel we estimate $x_p$ by $\hat x_p$.
 This estimator is Type~8 of quantile estimators on the software package R, \cite{R}.
  Given such a single sample,
 and fixed  $0<p,q<1$ we estimate the ratio $\rho =\rho(p,q)=x_p/x_q$ by $\hat \rho =\widehat x_p/\widehat x_q$.

 Assuming $F$ has a  positive and continuous derivative $f=F'$ on its support, the derivative of the
quantile function $G=F^{-1}$ is given by $G'(p)=g(p)=1/f(x_p)$; this is the {\em quantile density} of \cite{ par-1979}, earlier called the {\em sparsity index} by \cite{tukey-1965}. It arises in first order asymptotic covariance expressions,
see \cite[Ch.2]{david-1981} or \cite[Ch.7]{Das-2008}, where it is shown
that as $n$ increases without bound  $\e (\hat x_p)\doteq x_p$; and, for $0<p<q<1$
 \begin{eqnarray}\label{covxpxq}
\nonumber
  n\,\var (\widehat x_p)&\doteq &p(1-p)g^2(p)\equiv \sigma^2_p \\
  n\,\cov (\widehat x_p,\widehat x_q)&\doteq & p(1-q)g(p)g(q)\equiv \sigma_{p,q}~,
\end{eqnarray}
where $\doteq $ means that lower order terms are omitted.  For the case $0<q<p<1$, $ n\,\cov (\widehat x_p,\widehat x_q)\doteq q(1-p)g(p)g(q)\equiv \sigma _{p,q}$.

It follows that for $0<p<q<1$ a first-order approximation to the correlation between $\widehat x_p$, $\widehat x_q$ is  $\corr (\widehat x_p,\widehat x_q)
\doteq \sigma_{p,q}/(\sigma _p\sigma _q)=\sqrt {p(1-q)/\{q(1-p)\}}\;>0$; for $0<q<p<1$ it is $\sqrt {q(1-p)/\{p(1-q)\}}\;>0.$ This asymptotic correlation
 is notably free of $F$
and sample sizes, and must be taken into account in computing standard errors of $\widehat \rho =\widehat x_p/\widehat x_q.$
The classical formula \cite[p.50]{J-K-K-1993} for the variance of a ratio of random variables, provided the denominator has positive support, is given in
terms of means, variances and covariance of its components. We only consider $F$ with positive support, and thus when applied to sample quantile estimators with $0<p,q<1$ this formula for the ratio of variances
can be written:
\begin{equation}\label{asymvarrhohat}
    n\var (\widehat \rho _{p,q})   \doteq  n\var (\widehat x_p- \rho \,\widehat x_{q})/x_q^2\doteq a_0 +a_1\rho+a_2\rho^2\equiv h^2(\rho )~,
\end{equation}
where $h^2(\rho ) $ is the quadratic with constants defined in terms of (\ref{covxpxq}) by
$ a_0 =\sigma ^2_p/x_q^2$, $a_1=-2\sigma_{p,q}/x_q^2$ and $a_2=\sigma ^2_q/x_q^2.$
Note that $a_0,$ $a_1$ and $a_2$ are free of scale and sample size. The quadratic
$h^2(\rho)>0$ for all $\rho $  because $a_0>0$ and its discriminant $a_1^2-4a_0a_2<0$; the latter inequality follows from $ \corr (\widehat x_p,\widehat x_q) < 1.$

These results suggest that the large sample variance of $\widehat \rho =\widehat \rho _{p,q}$ is approximately
$h^2(\widehat \rho)/n$, and, because the large sample squared bias is of smaller order, the  standard error can be approximated by $\se (\widehat \rho )\doteq h(\widehat \rho )/\sqrt n\,.$
This formula has been derived for {\em known} constants $a_0$, $a_1$ and $a_2$. To make it distribution-free, one needs to replace $x_q$ by $\widehat x_q$, $g(p)$ by $\widehat g(p)$ and
$g(q)$ by $\widehat g(q)$; where $\widehat g$ is a quantile density estimate such
as the kernel density estimator described in Appendix~\ref{sec:qden}.   When this is done, we obtain the distribution-free standard error estimate
$\widehat {\se  }(\widehat \rho)=\widehat h(\widehat \rho)/\sqrt n\,.$

\subsection{Two interval estimators for ratios of quantiles}\label{subsec:DFcis}

We compare distribution-free large-sample confidence intervals for $\rho =x_p/x_q$, where $0<p,q<1$. The distribution of $n^{1/2}(\widehat \rho-\rho )$ is asymptotically normal but quite skewed for moderate sample sizes, so transformations
are employed to normalize its distribution and derive confidence intervals. This
methodology is standard, so here we only
present the final results for the log-transformation and a variance stabilizing transformation, with details given in
Section~\ref{sec:taylor}.

\subsubsection*{Studentized log-transformed ratios.}
 One traditional approach for finding confidence intervals of a ratio of statistics such as $\rho =\rho _{p,q}$  is to first find approximations to the asymptotic mean and variance of the estimated log-ratio $\widehat \theta = \ln (\widehat \rho )$, use the studentized version of this estimator to obtain a confidence interval for $\theta $ and then exponentiate this interval.
In terms of earlier definitions (\ref{covxpxq}), the asymptotic mean and variance of  $\widehat \theta =\ln (\widehat \rho)$ are shown in Appendix~\ref{sec:taylor} to be:
\begin{eqnarray} \label{asymmeanthetahat}
\e (\widehat \theta )&\doteq & \theta +\frac{1}{2n}\left (\frac {\sigma_q^2}{x_q^2}-\frac{\sigma ^2_p}{x^2_p}\right) \\ \label{asymvarthetahat}
\var ( \widehat \theta ) &\doteq &  \frac{1}{n}\left (\frac {\sigma ^2_p}{x_p^2}+
\frac {\sigma ^2_q}{x_q^2}-\frac {2\sigma _{p,q}}{x_px_q}\right )= \frac{1}{n}\;\frac{h^2(\rho )}{\rho ^2}~,
\end{eqnarray}
where $h^2(\rho)$ is given by (\ref{asymvarrhohat}).
The asymptotic normality of $\widehat \theta $ then leads to the nominal 100$(1-\alpha)$\% confidence interval for $\rho $:
\begin{equation}\label{Scis}
    [L,U]_\text{S} = \widehat \rho \,\left \{\exp \left [\mp z_{1-\alpha /2}\,\sqrt {{\var }( \widehat \theta )} \;\right ]\right \}~.
\end{equation}
To make the intervals (\ref{Scis}) distribution-free, the values of $x_p$, $x_q$, $\sigma _p$, $\sigma _q$ and $\sigma _{p,q}$ appearing in $\var ( \widehat \theta )$ need to be consistently estimated, and $\var ( \widehat \theta )$ replaced by $\widehat {\var }( \widehat \theta )$.
 It is also noted in Appendix~\ref{sec:qden}
 that the widths of these intervals behave like:
\begin{equation}\label{widthScis}
W_\text{S}\doteq 2\,\rho \,z_{1-\alpha /2}\,\sqrt {\var ( \widehat \theta )}\;=\,\frac {2\,z_{1-\alpha /2}h(\rho )}{\sqrt n}~.
\end{equation}

\subsubsection*{Variance stabilized ratios of quantiles.}
 Let $l(\rho )=a_1+2a_2\rho $ be the derivative
of the quadratic $h^2(\rho )=a_0+a_1\rho +a_2\rho ^2$ defined in (\ref{asymvarrhohat}), and let $D^2=4a_0a_2-a_1^2$ be the negative of its discriminant. Then as explained in Appendix~\ref{sec:qden}, one can derive large-sample nominal 100$(1-\alpha)$\% confidence intervals:
\begin{equation}\label{VSTcis}
    [L,U]_\text{V}=\frac {1}{2\widehat a_2}\left \{ \widehat D\sinh\left[\sinh^{-1}\left(\frac {\widehat l(\widehat \rho )}{\widehat D}\right ) \mp z_{1-\alpha /2}\;\sqrt {\frac {\widehat a_2}{n}}\, \right ]-\widehat a_1\right \}~.
\end{equation}
The hats appearing on $\widehat a_2$, $\widehat l$ and $\widehat D$ indicate that they are
the result of distribution-free estimates of $x_p,$ $\sigma _p$, $\sigma _{p,q}$ etc.
being replaced by consistent estimates.

The asymptotic widths of these intervals are, up to first order, the same
as those derived by the log-transformation (\ref{widthScis}).
Thus the large-sample coverage
and widths of the two intervals $[L,U]_\text{S}$ and $[L,U]_\text{V}$ are the same; so in Section~\ref{sec:simulations} we  compare their finite sample properties.

\section{Simulation Studies}\label{sec:simulations}

In this section we report simulated coverage probabilities and mean interval widths for several distributions.  Extensive simulations were carried out for the LN(0,1), EXP(1), $\chi^2_1$, $\chi^2_3$, $\chi^2_5$, Pareto(1), Pareto(1.5) and Pareto(2) distributions.  By \lq Pareto($a$)\rq\ we mean the Type II Pareto distribution with shape parameter $a$ and distribution function
$F_a(x)=1-(1+x)^{-a},$ for $a,x>0.$  Commands for generating data or finding quantiles from this distribution are
obtained by downloading the package {\tt actuar} on R.
We report the results for three of these distributions and remark that similar results were obtained for the other distributions.

\subsection{Moderate sample sizes}\label{moderaten}

\marginpar{Table 3 here.}

To ensure that the interval widths are considered in the correct context, in Table \ref{table3} we provide the true quantile ratios for the distributions considered.
In Table \ref{table4} we report the simulated coverage probabilities (cp) and average widths ($\overline{w}$) for the interval estimators associated with the LN(0,1), $\chi^2_3$ and the Pareto(2) distributions for various choices or $p$ and $q$ and three sample sizes $n=100, 250$ and 500.  In almost all cases, the coverage probabilities are between 0.95 and 0.97.  In general, the VST intervals are slightly narrower than the studentized intervals and consequently slightly less conservative. This is consistent with the folkloric view amongst applied statisticians that variance stabilization generally leads to more powerful tests than studentization for moderate sample sizes; a view recently reinforced by examples in \cite{KMS-2010}, \cite{S-2014} and theory in \cite{M-S-2012}.

\marginpar{Table 4 here.}
The results of Table~\ref{table4} were restricted to the special case of $q=1-p$ for choices of $p=0.05, 0.1, 0.2, 0.8, 0.9, 0.95$.  However, it will be useful to consider the coverage probabilities for a much wider choice of $p$ and $q$.  We will now consider coverage probabilities for the VST and studentized intervals for two of the distributions.  Additionally, we use the log-normal QOR only since the smaller computation cost means that we can use a large number of iterations over many choices of $p$ and $q$.

\marginpar{Figure 2 here.}

 As in \cite{prst-2014}, we use contour plots  to assess coverage probability over a wider range of $p,q$ combinations.  In Figure \ref{fig2} we plot the simulated coverage probabilities based on 10,000
 replications for all combinations of $p$ and $q$ from $0.05, 0.06,\dots,0.95$ for data sampled from the LOGN(0, 1) distribution.  Green indicates ideal coverage of between 0.95 and 0.96 (e.g. at least nominal) and light blue indicates slightly conservative intervals.  When $n=100$ we can see that the intervals can be very conservative (i.e. the dark blue regions) when $p$ and $q$ are close together.  However, such choices of $p$ and $q$ do not typically provide much insight since quantiles are approximately the same.  For other choices of $p$ and $q$ the coverages are quite good, despite the small sample sizes of $n=100$.  Typically, the VST interval is the marginally better performer with coverages slightly closer to the nominal level of 0.95.  As $n$ is increased to 250 and then to 500, we see that the coverage probabilities become even closer to nominal with a tendency for slightly conservative intervals.  Very rarely does the simulated coverage fall below the nominal coverage of 0.95 highlighting reliable performance for this distribution.

\marginpar{Figure 3 here.}

In Figure \ref{fig3} are shown the simulated coverage probability contour plots for the Pareto(2) distribution.  In general, the intervals are slightly more conservative than they were for the log normal although lower than nominal coverage is very rare.  Again, $p\approx q$ results in the most conservative intervals, especially for smaller $n$, although in practice this scenario is trivial, at best.  Coverage improves for increasing $n$ with most reported coverages between 0.95 and 0.97 when $n=500$.  In the next section we will see that further increases of the samples sizes continues to improve coverage.

\subsection{Large sample sizes}\label{largen}

In Table \ref{table5} we report large sample size empirical probabilities and mean widths for the same intervals and distributions summarized in Table~\ref{table4}, and one can see that the coverage probabilities are closer to nominal.   Further, the widths of the intervals are not so small as to justify the use of point estimates only.  This is especially true for the Pareto(2) distribution where even for $n=10,000$ the mean widths  are still large relative to the ratio being estimated; (e.g., $p=0.9$, $q=0.1$ with $\rho = 39.97$ and the mean interval width is 5.89 for both intervals).

\marginpar{Table 5 here.}

\section{Effects of contamination}\label{sec:contamination}

\subsection{Mixture distribution with spike at zero}\label{sec:simmix}

Additional to the large sample simulations conducted above, we note that many samples of income data include a small percentage of zero values (e.g. for households with zero income or households in debt rounded upwards to zero). We therefore examine the following mixture model:

\begin{equation}
F_\epsilon = (1-\epsilon)F + \epsilon\Delta_0\label{Feps}
\end{equation}
where $F$ is the positive income distribution , $\Delta_0$ is places all its mass at the point 0 and $0< \epsilon < 1$ is the proportion of the mixture that are zeroes.

\marginpar{Table 6 here.}

In Table \ref{table6} we report simulation results for zeroes mixed with the  LOGN(0,1), $\chi^2_3$ and the Pareto(2) distributions, respectively, with probabilities $(\epsilon ,1-\epsilon)$.  For simplicity we report only the coverage probabilities and only for the VST intervals; similar results are obtained for the studentized intervals.  When $\epsilon=0.05$ we do not report results for any ratio estimating $x_{0.05}$ since approximately half of the estimates will equate to zero.  Overall the coverage probabilities are close to nominal with a tendency for conservative intervals when estimating $x_{0.05}$.  In this case a mass of zeroes lying close to one of the quantiles in the ratio will have a small effect on the estimated density in that vicinity.

\subsection{Robustness properties}\label{sec:IF}

 For background material on robustness concepts such as influence functions and breakdown points, see \cite{HRRS86} or \cite{S-S-1990}.

\subsubsection*{Influence functions.}
Extending on the zero mixture distribution from \eqref{Feps}, define the \lq contamination\rq\  distribution which places positive  probability $\epsilon $ on $z$ (the contamination point) and $1-\epsilon $ on the distribution $F$. Formally, it is defined for each $x$ by $F_\epsilon ^{(z)}(x)\equiv (1-\epsilon)F(x)+\epsilon I[x\geq z]$.   The influence function for any functional $T(F)$ is then defined for each $z$ as
 the $\IF (z;T,F)\equiv \lim _{\epsilon \to 0}\frac {\partial }{\partial \epsilon }T( F_\epsilon ^{(z)})$ \citep[see][]{HA74}.
 The influence function of the $p$th quantile $x_p=G(F;p)=F^{-1}(p)$ is well-known \cite[p.59]{S-S-1990} to be
\begin{equation}\label{IFquantile}
    \IF (z;\,G(\,\cdot )\, ,p),F)= \{p-I[x_p\geq z]\}\,g(p)~,
\end{equation}
where $G'(p)=g(p)= 1/f(x_p)$ is the {\em quantile density} of $G$ at $p$.
One can show that $\e _F [\IF (Z;\,G(\,\cdot )\, ,p),F),F)]=0$ and $\var _F [\IF (Z;\,G(\,\cdot )\, ,p),F),F)]=\e _F [\IF ^2(Z;\,G(\,\cdot\, ,p),F),F)]=p(1-p)\,g^2(p)$. The reason for calculating this variance
is that it arises in the asymptotic variance of the functional applied to the empirical distribution $F_n$, namely $G(F_n ,p)$; that is, $n\;\var [G(F_n ,p)]=p(1-p)\,g^2(p)$; and sometimes a simple expression for the asymptotic variance is not otherwise available.

The influence function of the ratio of two quantiles $\rho _{p,q}(F)=x_p/x_q=G(\,\cdot )\, ,p)/G(\,\cdot)\, ,q)$ is then by elementary calculus and (\ref{IFquantile}) found to be
\begin{eqnarray}\label{IFL1}
  \IF (z;\,\rho _{p,q} ,F) &=& \frac {\IF (z;\,G(\,\cdot )\, ,p),F)}{x_{q}}
         -\frac {x_{p}\IF (z;\,G(\,\cdot )\, ,q),F)}{x_{q}^2}\\ \nonumber
   &=&  \frac{x_{q}\{p-I[x_{p}\geq z]\}\,g(p)-
                  x_{p}\{q-I[x_{q}\geq z]\}\,g(q)}{x_{q}^2}~.
 \end{eqnarray}
When expanded in a power series expansion with respect to $\epsilon$, we have that $\rho _{p,q}(F_\epsilon ^{(z)})= \rho _{p,q}(F) + \epsilon \IF (z;\,\rho _{p,q} ,F) + O(\epsilon^2)$.  Consequently, it would be of interest to study the influence relative to $\rho _{p,q}(F)$ since large values of $\IF (z;\,\rho _{p,q} ,F)$ are not suggestive of high sensitivity if the ratio at $F$ is very large.

\marginpar{Figure 4 here.}

To assess influence sensitivity relative to the size of the ratio at $F$, in Plot A of Figure \ref{fig4} is shown $\IF (z;\,\rho _{p,q} ,F)/\rho _{p,q}(F)$ for $z\in [0,1]$ and $p \in (0.05, 0.95)$ and $q=1-p$.  As one can see, the influence increases quickly as $p$ approaches its boundaries
when $z$ is close to zero. In this situation either $x_p$ or $x_q$ is close to zero and therefore close to the contamination.  In Plot B we vary both $p$ and $q$ but fix the contamination $z=0$.  Again it can be seen that the ratio estimator is especially sensitive to zero valued observations when either $p$ or $q$ is close to 0.   In practice, if a data set contains a mixture of zero valued observations together with positive values then inference will be difficult if either $p$ or $q$ is small.  In Section \ref{sec:simmix} simulations revealed that even a small proportion of zeroes could result in over conservative intervals when $p$ or $q$ was equal to 0.05.

\marginpar{Figure 5 here.}

The influence function can also be used to calculate the asymptotic variance $$n\;\var [\rho_{p,q}(F_n)-\rho_{p,q}(F)]=\text{ASV}(\rho_{p,q};F)=\e\left[\IF (z;\,\rho _{p,q} ,F)^2\right]$$ by expanding \eqref{IFL1} and noting that for the two cases $p < q$ and $p>q$ we have $I(x_{p}\geq z])I(x_{q}\geq z])=I(x_{p}\geq z])$ and $I(x_{p}\geq z])I(x_{q}\geq z])=I(x_{q}\geq z])$ respectively.  This gives
\begin{equation}
\text{ASV}(\rho_{p,q};F) = \frac{1}{x_q^4}\left[p(1-p)x_q^2g^2(p)+q(1-q)x_p^2g^2(q)-2x_qx_pm(p,q)g(p)g(q)\right]\label{ASV}
\end{equation}
where $m(p,q)=p(1-q)$ when $p<q$ and $q(1-p)$ when $p > q$.  It can be verified that this expression for the asymptotic variance is equal to \eqref{asymvarrhohat}.  Also, for the special case $p=q$ we have simply
\begin{equation}
\text{ASV}(\rho_{p,1-p};F) = \frac{p(1-p)}{x_{1-p}^4}\left[x_{1-p}g(p)-x_pg(1-p)\right]^2\label{ASV2}
\end{equation}
We assess the variability of the ratio estimator with respect to the magnitude of the ratio to be estimated.  Therefore, in Figure \ref{fig5} we plot $\text{ASV}(\rho_{p,1-p};F)/\rho^2_{p,1-p}(F)$ for $p$ in $(0.05, 0.95)$ (Plot B).  These plots show that the variance of the ratio estimator can be very large (relative the population ratio
squared) when either $p$ and $q$ is close to zero.  In practice, one needs to be aware that ratios involving very small quantiles will have higher variability and wider intervals relative to the magnitude of the ratio will result.

\subsubsection*{Breakdown points.}

The asymptotic breakdown point $\epsilon ^*=\epsilon ^*(T,F)$ of a functional $T(F)$ is roughly speaking the minimum proportion of contamination of $F$ to $F_\epsilon ^{(z)}$ that can render useless $T(F_\epsilon ^{(z)})$, as $z$ varies over the support of $F$. This $\epsilon ^*(T,F)$ is often free of $F$ and gives an indication of how sensitive the functional $T(F)$ and its estimator $T(F_n)$ are to possible contamination. Unfortunately, rigorous definitions and mathematical arguments for finding such
breakdown points are often complicated, see \cite{gent-2003} and references therein.
Here we give a somewhat heuristic derivation of the breakdown point for the functional $T_{p,q}(F)=\rho _{p,q}=T_p(F)/T_q(F),$ where $T_p(F)=F^{-1}(p)$ and $F$ is continuous and strictly increasing on $(0,\infty )$.

It is well known and intuitively clear that the breakdown point
of $T_p(F)$ itself is $\epsilon ^*(T_p,F)=\min \{p,1-p\}.$ This is because if $\epsilon \geq p$ one can move the $p$th quantile
of  $T_p(F_\epsilon ^{(z)})$   to 0 by choice of $z$ and if $\epsilon >1-p$ one can make it move towards $+\infty .$  And for any $\epsilon < \min \{p,1-p\} $ the contamination cannot move $T_p(F_\epsilon ^{(z)})$ to one of its boundaries.

The functional $T_{p,q}(F)=\rho _{p,q}$ is more complicated, and \lq breaks down\rq\ if either $T_p$ or $T_q$ breaks down,
(because then the ratio is $0,+\infty $ or undefined), and hence uninformative.  It also  breaks down if $T_p(F _\epsilon ^{(z)})= T_q(F_\epsilon ^{(z)})$  (because then the ratio is 1, another uninformative value); and this can be arranged if
and only if $\epsilon \geq |p-q|$ by taking $z=x_p$.
Putting these facts together, the  breakdown point for the ratio of quantiles  equals $\epsilon ^*(T_{p,q})=  \min \{\{p,1-p,q,1-q,|p-q|\}>0$. This breakdown point is clearly maximized by taking $p=1/3,q=2/3$ or $p=2/3,q=1/3.$

\section{Discussion and further research}\label{sec:summary}

 While point estimators of the ratio of percentiles from a single sample are commonplace, accompanying standard errors and/or interval estimators of such ratios are now possible. We have shown that such procedures are necessary because what are  usually considered large samples do not by any means guarantee that variability is negligible in the ratio estimates.

  We compared two interval estimators of the quantile ratios, one based on the
  studentized log-transformation, and the other on variance stabilization. While asymptotically equivalent, simulations showed that the coverage of the VST intervals was slightly better than the log intervals, although both
  are somewhat conservative for moderate sample sizes. However, the log-transformed  ratios are more amenable to
  computing two-sample tests from independent samples, as described in \ref{sec:twotest}.

  One may be able to reduce the conservative coverage of both intervals by using a bias correction; for example
  by subtracting an estimate of the bias in $\log (\hat \rho)$, see Equation (\ref{asymmeanthetahat}). However,
  we tried this and other bias correction methods for the variance-stabilized estimate ratio, to no avail.
   Finally, it may well be possible to choose sample sizes to achieve a desired relative width in the confidence intervals over a  large class of distributions.

   The good robustness properties of simple ratios of quantiles are desirable  in
   all inequality measures; and, no doubt replacing moments by appropriate quantiles in more sophisticated inequality measures is possible and another area of further research.


\section{Appendix}\label{sec:appendix}

\subsection{ABS Data}\label{sec:ABSdata}

\marginpar{Table 7 here.}

  The number of persons (in thousands) for each income category is listed: for example, there were
73,700 persons with no income in the financial year beginning 1 July, 2005.  The total number of persons in this year is estimated at 19,930,700.   Of course not all households were sampled and converted to
equivalized disposal income per person. On page 25 of the same ABS document one
finds that the sample size of households was 9,961 for 2005 and 14,569 for 2011.
Thus the figures in Table \ref{table7} are only estimates based on what was found in
the samples, and then converted to population estimates.

The original sample equivalized data are not readily available, so we
 \lq reconstructed\rq\ the sample by generating random numbers within each
 income range in proportion to those in Table \ref{table7}. The ABS informs us that different
 weights for each income group were used to generate the table, and these are confidential for
 privacy reasons, so our reconstructed sample will differ from theirs; nevertheless we think
 the differences are negligible for our purposes.
        Also, we will truncate the income data to the interval $(0,2000]$ for two reasons: first,
         to obtain a sample from a continuous data set by excluding the positive mass at 0; and second because the largest category \lq 2000 or more\rq\ is unbounded.  Our reconstructed sample for 2005 has size  $9961(1-(73.7+506.2)/19930.7)=9671$ and similarly for 2011 it is 13904.

\subsection{Derivation of (\ref{asymmeanthetahat})-(\ref{widthScis})} \label{sec:taylor}

In what follows, we use the general approximations derived from Taylor expansions
$\e [\ln(U)] \doteq \ln (\e[U])-\var [U]/\{2\e ^2[U]\}$ and $\var [\ln(U)] \doteq \var [U]/\{\e ^2[U]\}$ and similarly for $V$.
Further, we need
\begin{equation}\nonumber
    \e [\ln (U)\,\ln (V)]\doteq \ln (\e[U])\,\ln (\e[V])+\frac {\cov[U,V]}{\e[U]\,\e[V]}
-\frac {\ln (\e[V])\,\var [U]}{2\e^2[U]} -\frac {\ln (\e[U])\,\var [V]}{2\e^2[V]}~.
\end{equation}
Combining the above formulae, the approximate variance of $\ln (U/V)$ is
\begin{equation}\nonumber
    \var \left [\ln \left (\frac {U}{V}\right )\right ]\doteq \frac {\var [U]} {\e^2[U]}+ \frac {\var [V]} {\e^2[V]}-\frac {2\cov [U,V]}{\e[U]\,\e[V]} ~.
\end{equation}
Applying these approximations to $U=\widehat x_p$ and $V=\widehat x_q$ yields
(\ref{asymmeanthetahat}) and (\ref{asymvarthetahat}).
  By the Delta Theorem, \cite[p.40]{Das-2008}, $(\widehat \theta -\theta )/\sqrt {\var [ \widehat \theta ]}$ converges in distribution to a standard normal distribution, so a large sample 100$(1-\alpha )$\% confidence interval for $\theta $ is
given by $\widehat \theta \mp z_{1-\alpha /2}\,\sqrt {\var [ \widehat \theta ]}\;.$ This leads immediately to the interval (\ref{Scis}) having
 the same confidence for $\rho _{p,q}$.
By expanding the exponentials appearing in  (\ref{Scis}) in series, it is found that the widths of these intervals $W_\text{S}=U_\text{S}-L_\text{S}$ can be expressed in terms of $\rho $ and $\var [\widehat \theta ]$ as shown in (\ref{widthScis}).

\subsection{Quantile density estimation}\label{sec:qden}

The confidence intervals described previously (\ref{Scis}) and (\ref{VSTcis}) require estimates of $a_0$, $a_1$  and
$a_2$ appearing in the asymptotic variance quadratic (\ref{asymvarrhohat}), which require estimates of $\sigma _p$, $\sigma _q$ and $\sigma _{p,q}$ defined in (\ref{covxpxq}); and these in turn require estimates of the quantile densities
$g(p)$ and $g(q)$.  There have been many contributors to this problem and we
refer the reader to  \cite{ps-2014} for background and results on kernel density
estimators of the form $\widehat g(p)=\sum _{i=1}^nX_{(i)}\,\{k_b(p-\frac {(i-1)}{n})-k_b(p-\frac {i}{n} ) \},$
where $b$ is a bandwidth and $k_b(\cdot)=k(\cdot -b)/b$ for some kernel function $k$ which  is an even function
on $[-1,1]$. We follow \cite{ps-2014} in using the Epanechnikov kernel with an estimated optimal bandwidth. The optimal bandwidth depends on the quantile optimality ratio QOR$(u)=g/g''(u)$ and the QOR for an assumed underlying log-normal distribution can be used for many unimodal distributions supported on the half-infinite interval $[0,\infty)$;  a boundary correction is included for quantiles near 0.  Alternatively, one can calculate the QOR assuming that the underlying density can be well-approximated by the highly flexible generalized lambda distribution (GLD), see Section~\ref{sec:GLD}.

The intervals (\ref{VSTcis}) are derived exactly as  for the quantile-based skewness coefficients in \cite[Sec. 3.3]{S-2014} and displayed in Equation~9 of that paper. One only needs to replace the
coefficients in the quadratic defining the asymptotic variance by the simpler ones needed here (\ref{asymvarrhohat}).
It is also shown there that the width $W_\text{V} =U_\text{V}-L_\text{V}$ can
be expressed $ W_\text{V}= 2\,\sqrt {g (\rho )} \;z_{1-\alpha /2}/\sqrt n\,   +o_p(n^{-1/2})$. The leading term of this expression is exactly equal to that in (\ref{widthScis}), which is
the asymptotic width for the interval $W_\text{S}$ based on studentization.

\subsection{GLD methods and results}\label{sec:GLD}

GLD QOR identifies another approach when the underlying distribution is assumed to be at least close to a member of the highly-flexible generalized lambda distribution.  For more on the estimation of the quantile density see Appendix \ref{sec:qden} and \cite{ps-2014}. In general, the VST intervals are slightly narrower than the studentized intervals and consequently slightly less conservative.  Additionally, there may be some small gain to using the GLD QOR, in particular when the distribution is not the log-normal.  However, the log-normal QOR provides a good bandwidth and is easier to compute.  Given that there were 10,000 iterations used in the simulations, we used method of moments estimators for the GLD distribution which were comparatively quick to compute.

\marginpar{Table 8 here.}

There are various other GLD estimators available \citep[for a recent discussion see ][]{CO&ME15}.  The R packages \texttt{gld} \citep{KI&DE14} and \texttt{GLDEX} \citep{SU07R} provide various GLD estimators.
 However, some small improvements may results when using the GLD QOR as seen in Table \ref{table8}.
  Using the parameterisation of \cite[][FKML parameterisation]{fmkl-1988}, some small improvements may be achieved although requiring the estimation of four parameters increases the computational complexity.

\subsection{Intervals comparing two independent ratios }\label{sec:twotest}

The theory supporting the studentized log-transformed ratios can also be extended to consider the difference between two independent log-transformed ratio estimators.  For simplicity we will assume that the same $p$ and $q$ are used for each of the estimators although this is technically not required.  Let $\widehat{\rho}_x = \widehat{\rho}_x(p,q)$ and  $\widehat{\rho}_y = \widehat{\rho}_y(p,q)$ be estimates of the percentile ratios $\rho_x$ and $\rho_y$ respectively.  Further, let $\widehat{\theta}_x=\ln(\widehat{\rho}_x)$ and $\widehat{\theta}_y=\ln(\widehat{\rho}_y)$ where the asymptotic variances, $\var ( \widehat \theta_y )$ and $\var ( \widehat \theta_y )$, for each can be obtained from \eqref{asymvarthetahat}.  Then a large sample $100(1-\alpha)$\% confidence interval for $\ln(\rho_x)-\ln(\rho_y)$ is
\begin{equation}
(\widehat{\theta}_x - \widehat{\theta}_y) \pm z_{1-\alpha/2}\sqrt{\var ( \widehat \theta_x ) + \var ( \widehat \theta_y )}\label{ci_diff}
\end{equation}
or, for $\rho_x/\rho_y$ to be interpreted on a ratio scale,
\begin{equation}
\frac{\widehat{\rho}_x}{\widehat{\rho}_y} \left \{\exp \left [\mp z_{1-\alpha /2}\,\sqrt {{\var }( \widehat \theta_x) + {\var }( \widehat \theta_y)} \;\right ]\right \}\label{ci_diff2}.
\end{equation}

The good empirical coverage probabilities for the interval estimates of a single ratio suggest good approximations for the standard error which in turn suggest good coverage is achievable when considering two independent ratios.  We provide some brief verification here via simulation and note that these coverage probability results are for both of the interval estimators in \eqref{ci_diff} and \eqref{ci_diff2} which are equivalent in this regard.

\marginpar{Table 9 here.}

Empirical coverage probabilities computed over 10,000 simulation runs are reported in Table \ref{table9}.  The samples sizes were $n$ and $m$ for each of the two groups with data sampled from the LN(0,1) and LN(0.2,1.5) distributions respectively.  While slightly conservative, for each of the differences in percentile ratios considered for this simulation, the coverage does not drop below the nominal level of 0.95.  Additionally, improved coverage is observed for increasing sample sizes.

\clearpage
\newpage

\subsection{R script for computing confidence intervals}\label{Rscript}
\renewcommand{\baselinestretch}{1.25}

\begin{small}
\begin{verbatim}

############# R script by Luke A. Prendergast, 28 August, 2015

Epanechnikov <- function(u){
  3*(1 - u^2)*(abs(u) <= 1)/4}

QuantileDensity <- function(x, p, correct = TRUE){
  # This function computes the quantile density associated with
  # the p-th quantile.  The Epanechnikov kernal density estimator
  # is used with an optimal bandwidth selected based on the QOR
  # for the LNORM distribution.
  #
  # Args:
  #   x: A numeric vector.
  #   p: A numeric value between 0 and 1.
  #   correct: If correct = TRUE then a boundary correction will
  #            be carried out if p is less than the bandwidth.

  # Compute the QOR for the LNORM distribution.
  qPhiu <- 1/dnorm(qnorm(p))
  qPhipru <- qnorm(p)*qPhiu^2
  qPhiprpru <- (qPhiu^3)*(1 + 2*qnorm(p)^2)
  QLNu <- qlnorm(p)
  qLNu <- QLNu*qPhiu
  qLNpru <- qLNu*qPhiu + QLNu*qPhipru
  qLNprpru <- qLNpru*qPhiu + 2*qLNu*qPhipru + QLNu*qPhiprpru
  qratio <- qLNu/qLNprpru

  n <- length(x)
  bw <- (15^(1/5))*(qratio)^(2/5)/(n^(1/5))
  if (correct) bw <- min(p, bw)

  xsort <- sort(x)
  consts <- (Epanechnikov((p - (1:n - 1)/n)/bw)
             - Epanechnikov((p - (1:n)/n)/bw))/bw
  return(sum(xsort*consts))
  }


ratioCI <- function(x, p, q, conf.level = 0.95, correct = TRUE)
{
  # This function computes the studentised and VST confidence
  # intervals for the ratio of the p-th to q-th quantiles.
  #
  # Args:
  #   x: A numeric vector.
  #   p: A numeric value between 0 and 1.
  #   q: A numeric value between 0 and 1.
  #   conf.level: A numeric value between 0 and 1 specifying
  #               the coverage probability for the intervals.
  #   correct: Choice to carry out boundary correction passed
  #            to QuantileDensity.

  zcrit <- qnorm(1 - (1 - conf.level)/2)
  n <- length(x)
  Ghat <- quantile(x, c(p, q), type = 8, names = FALSE)
  xphat <- Ghat[1]
  xqhat <- Ghat[2]
  rhopqhat <- xphat/xqhat

  gphat <- QuantileDensity(x, p, correct = TRUE)
  gqhat <- QuantileDensity(x, q, correct = TRUE)

  mpq <- min(p, q)
  Mpq <- max(p, q)

  # The VST interval
  a0hat <- (p*(1 - p)*gphat^2)/xqhat^2
  a1hat <-  -2*mpq*(1 - Mpq)*gphat*gqhat/xqhat^2
  a2hat <-  (q*(1 - q)*gqhat^2)/xqhat^2

  hsqhat <- a0hat + a1hat*rhopqhat + a2hat*rhopqhat^2
  lhat <- a1hat + 2*a2hat*rhopqhat
  asymSErhopqhat <- sqrt(hsqhat/n)

  Dhat <- sqrt(4*a0hat*a2hat - a1hat^2)
  chat <- zcrit*sqrt(a2hat/n)
  CI.vst <- (Dhat*sinh(asinh(lhat/Dhat) + c(-1, 1)*chat) - a1hat)/(2*a2hat)

  # The studentized interval
  nvarthetahat <- p*(1 - p)*gphat^2/xphat^2 +
    q*(1 - q)*gqhat^2/xqhat^2 - 2*mpq*(1 - Mpq)*gphat*gqhat/(xphat*xqhat)
  sigma_n <- sqrt(nvarthetahat/n)
  CI.stud <- rhopqhat*exp(c(-1, 1)*zcrit*sigma_n)

  CIs <- rbind(CI.vst, CI.stud)
  rownames(CIs) <- c("VST", "Stud")

return(list(rho.hat = rhopqhat, CIs = CIs))
}

##############################################################################

# An example for LNORM generated data

p <- 0.9
q <- 0.1

true.rho <- qlnorm(p)/qlnorm(q)
true.rho

x <- rlnorm(1000)
ratioCI(x, 0.9, 0.1)
\end{verbatim}
\end{small}

\clearpage
\newpage

 \begin{table}
 \centering
\caption{Ratios of EWI percentiles reported on page 25 of \cite{ABSincomedata} over selected years from 2003 to 2011.  PX/PY denote the ratio of the X-th percentile to the Y-th percentile.\label{table1}}
\vspace{.2cm}
\begin{tabular}{llllll}
  Ratio   & 2003 & 2005 & 2007 & 2009 & 2011  \\
  \hline
  P90/P10 & 3.87 & 4.05 & 4.35 & 4.24 & 4.10  \\
  P80/P20 & 2.55 & 2.58 & 2.60 & 2.70 & 2.61  \\
  P80/P50 & 1.53 & 1.55 & 1.58 & 1.60 & 1.56  \\
  P20/P50 & 0.60 & 0.60 & 0.59 & 0.59 & 0.60  \\
   \end{tabular}
\end{table}

\begin{table}
\centering
\begin{footnotesize}
\caption{Estimated ratios $\widehat{\rho}$ and distribution-free (DF) studentized-log and VST intervals (Stud CI and VST CI; see Section \ref{subsec:DFcis} for these interval estimators) for the
data depicted in Figure \ref{fig1}.   Also, empirical coverage probabilities cp; mean widths: $\overline{w}$ based on 10,000 simulation runs from the fitted gamma distributions used to overlay the densities in Figure \ref{fig1}.  The ratios for the fitted
 gamma are denoted $\tilde \rho$.\label{table2}}
\vspace{.5cm}
\begin{tabular}{lrrrrr}
                            \multicolumn{6}{l}{2005} \\
                         & &   90/10 & 80/20 & 80/50 & 20/50  \\
  \cline{2-6} \\
DF&  $\widehat{\rho}$ & 3.888 & 2.502 & 1.515 & 0.605            \\
&  Stud CI       &[3.81,3.97] &[2.46,2.55]&[1.50, 1.53]&[0.596, 0.614]\\
&  VST CI          &[3.81,3.97]&[2.46,2.55]&[1.50, 1.53]&[0.596, 0.614]     \\ \\
Fitted&  $\tilde \rho$        & 3.872   &  2.419  &   1.507  &   0.623   \\
Gamma &  VST  cp         & 0.954 & 0.952 & 0.952 & 0.952             \\
& $\overline{w}$ & 0.201 & 0.092 & 0.039 & 0.020  \\
&  Stud  cp      & 0.954 & 0.952 & 0.952 & 0.952          \\
& $\overline{w}$ & 0.201 & 0.093 & 0.039 & 0.020  \\ \\
                       \multicolumn{6}{l}{2011} \\
                    &   & 90/10&80/20&80/50&20/50 \\
\cline{2-6} \\
DF & $\widehat{\rho}$   & 3.766 & 2.530 & 1.535 & 0.606 \\
 & Stud CI          & [3.70,3.83] & [2.49,2.57] & [1.52, 1.55] & [0.599, 0.614] \\
 & VST CI             & [3.70,3.83] & [2.49,2.57] & [1.52, 1.55] & [0.599, 0.614] \\ \\
Fitted & $\tilde \rho$        & 3.678 & 2.34  & 1.485 & 0.635 \\
Gamma & VST  cp         & 0.954 & 0.956 & 0.949 & 0.956 \\
 & $\overline{w}$ & 0.152 & 0.072 & 0.031 & 0.016 \\
 & Stud  cp      & 0.954 & 0.956 & 0.949 & 0.955 \\
 & $\overline{w}$ & 0.152 & 0.072 & 0.031 & 0.016 \\
\end{tabular}
\end{footnotesize}
\end{table}

\begin{table}
\caption{Values of $\rho_{p,q}$ for the three distributions LN(0,1), $\chi^2_3$ and Pareto(2) for which the coverage probabilities and intervals widths are reported in Table \ref{table4}.\label{table3}}
\centering
\vspace{0.5cm}
\begin{tabular}{rrrrrrr}
  \hline
  & 5/95 & 10/90 & 20/80 & 80/20 & 90/10 & 95/5  \\
  \hline
        LN & 0.04 & 0.08 & 0.19 & 5.38 & 12.98 & 26.84 \\
    $\chi^2_3$ & 0.04 & 0.09 & 0.22 & 4.62 & 10.70 & 22.21 \\
    PAR & 0.01 & 0.03 & 0.10 & 10.47 & 39.97 & 133.66\\ \end{tabular}
\end{table}

\begin{table}
\centering
\begin{footnotesize}
\caption{Coverage probabilities (cp) and mean ($\overline{w}$) interval width for the VST and studentized intervals (Stud) based on the lognormal-QOR bandwidth.\label{table4}}
\vspace{0.5cm}

\begin{tabular}{lrllrrrrrr} $n$ & $F$ &    &  & 5/95 &10/90 & 20/80 & 80/20 & 90/10 & 95/5 \\
  \hline
100   &LN         & VST  & cp             &0.965 & 0.966 & 0.970 & 0.964 & 0.965 & 0.966 \\
      &           &      & $\overline{w}$ &0.057 & 0.083 & 0.147 & 4.318 & 14.354 & 41.977 \\
      &           & Stud & cp             &0.966 & 0.968 & 0.972 & 0.966 & 0.967 & 0.969 \\
      &           &      & $\overline{w}$ &0.059 & 0.085 & 0.148 & 4.365 & 14.743 & 45.015 \\
      &$\chi^2_3$ & VST  & cp             &0.947 & 0.956 & 0.963 & 0.964 & 0.959 & 0.952 \\
      &           &      & $\overline{w}$ &0.062 & 0.095 & 0.155 & 3.477 & 12.013 & 39.261 \\
      &           & Stud & cp             &0.950 & 0.961 & 0.965 & 0.965 & 0.960 & 0.956 \\
      &           &      & $\overline{w}$ &0.067 & 0.098 & 0.157 & 3.483 & 12.084 & 39.704 \\
      &PAR        & VST  & cp             &0.958 & 0.956 & 0.966 & 0.963 & 0.960 & 0.957 \\
      &           &      & $\overline{w}$ &0.067 & 0.045 & 0.112 & 13.222 & 83.943 & 611.063 \\
      &           & Stud & cp             &0.966 & 0.963 & 0.967 & 0.970 & 0.966 & 0.965 \\
      &           &      & $\overline{w}$ &0.074 & 0.048 & 0.115 & 13.417 & 88.059 & 2585.756 \\      \hline
250   &LN         & VST  & cp             &0.966 & 0.968 & 0.969 & 0.964 & 0.971 & 0.964 \\
      &           &      & $\overline{w}$ &0.032 & 0.049 & 0.088 & 2.582 & 8.359 & 23.123 \\
      &           & Stud & cp             &0.970 & 0.966 & 0.969 & 0.964 & 0.971 & 0.966 \\
      &           &      & $\overline{w}$ &0.032 & 0.050 & 0.089 & 2.591 & 8.428 & 23.520 \\
      &$\chi^2_3$ & VST  & cp             &0.959 & 0.961 & 0.962 & 0.962 & 0.960 & 0.958 \\
      &           &      & $\overline{w}$ &0.039 & 0.059 & 0.097 & 2.104 & 7.033 & 21.030 \\
      &           & Stud & cp             &0.964 & 0.962 & 0.964 & 0.961 & 0.962 & 0.960 \\
      &           &      & $\overline{w}$ &0.040 & 0.060 & 0.097 & 2.105 & 7.047 & 21.112 \\
      &PAR        & VST  & cp             &0.960 & 0.957 & 0.957 & 0.963 & 0.959 & 0.959 \\
      &           &      & $\overline{w}$ &0.011 & 0.025 & 0.066 & 7.511 & 43.094 & 225.703 \\
      &           & Stud & cp             &0.962 & 0.959 & 0.960 & 0.965 & 0.962 & 0.964 \\
      &           &      & $\overline{w}$ &0.012 & 0.026 & 0.067 & 7.545 & 43.664 & 233.251 \\  \hline
500   &LN         & VST  & cp             &0.969 & 0.963 & 0.961 & 0.963 & 0.964 & 0.970 \\
      &           &      & $\overline{w}$ &0.022 & 0.034 & 0.061 & 1.771 & 5.739 & 15.761 \\
      &           & Stud & cp             &0.970 & 0.964 & 0.963 & 0.963 & 0.964 & 0.970 \\
      &           &      & $\overline{w}$ &0.022 & 0.034 & 0.061 & 1.774 & 5.760 & 15.875 \\
      &$\chi^2_3$ & VST  & cp             &0.960 & 0.963 & 0.958 & 0.961 & 0.962 & 0.963 \\
      &           &      & $\overline{w}$ &0.028 & 0.041 & 0.068 & 1.454 & 4.848 & 14.231 \\
      &           & Stud & cp             &0.960 & 0.964 & 0.960 & 0.961 & 0.961 & 0.963 \\
      &           &      & $\overline{w}$ &0.028 & 0.042 & 0.068 & 1.455 & 4.853 & 14.256 \\
      &PAR        & VST  & cp             & 0.960 & 0.957 & 0.962 & 0.958 & 0.957 & 0.959 \\
      &           &      & $\overline{w}$ & 0.007 & 0.017 & 0.046 & 5.081 & 28.285 & 140.555 \\
      &           & Stud & cp             & 0.961 & 0.959 & 0.961 & 0.960 & 0.958 & 0.959 \\
      &           &      & $\overline{w}$ &0.008 & 0.017 & 0.046 & 5.090 & 28.446 & 142.368 \\     \hline
       \end{tabular}
\end{footnotesize}
\end{table}

\begin{table}
\centering
\begin{footnotesize}
\caption{Large sample empirical probabilities (cp) and average interval width ($\overline{w}$) for the VST and studentized intervals (Stud) using the lognormal QOR.\label{table5}}
\vspace{0.5cm}
\begin{tabular}{lrllrrrrrr} $n$ & $F$ &    &  & 5/95 &10/90 & 20/80 & 80/20 & 90/10 & 95/5 \\
  \hline
1000   &LN         & VST  & cp             &0.966 & 0.964 & 0.959 & 0.959 & 0.963 & 0.968 \\
      &           &      & $\overline{w}$ &0.015 & 0.023 & 0.042 & 1.223 & 3.928 & 10.727 \\
      &           & Stud & cp             &0.966 & 0.963 & 0.959 & 0.959 & 0.963 & 0.967 \\
      &           &      & $\overline{w}$ &0.015 & 0.023 & 0.042 & 1.224 & 3.935 & 10.761 \\
      &$\chi^2_3$ & VST  & cp             &0.955 & 0.958 & 0.958 & 0.956 & 0.955 & 0.962 \\
      &           &      & $\overline{w}$ &0.019 & 0.029 & 0.047 & 1.010 & 3.342 & 9.775 \\
      &           & Stud & cp             &0.957 & 0.958 & 0.957 & 0.956 & 0.955 & 0.962 \\
      &           &      & $\overline{w}$ &0.020 & 0.029 & 0.047 & 1.010 & 3.344 & 9.784 \\
      &PAR        & VST  & cp             & 0.954 & 0.955 & 0.956 & 0.956 & 0.956 & 0.958 \\
      &           &      & $\overline{w}$ &0.005 & 0.012 & 0.032 & 3.522 & 19.396 & 93.367 \\
      &           & Stud & cp             & 0.954 & 0.957 & 0.956 & 0.958 & 0.955 & 0.959 \\
      &           &      & $\overline{w}$ &0.005 & 0.012 & 0.032 & 3.525 & 19.446 & 93.893 \\    \hline
5000  &LN         & VST  & cp             &0.958 & 0.956 & 0.953 & 0.952 & 0.955 & 0.957 \\
      &           &      & $\overline{w}$ &0.006 & 0.010 & 0.018 & 0.534 & 1.690 & 4.512 \\
      &           & Stud & cp             &0.958 & 0.955 & 0.953 & 0.953 & 0.955 & 0.956 \\
      &           &      & $\overline{w}$ &0.006 & 0.010 & 0.018 & 0.534 & 1.691 & 4.515 \\
      &$\chi^2_3$ & VST  & cp             &0.957 & 0.954 & 0.948 & 0.950 & 0.957 & 0.957 \\
      &           &      & $\overline{w}$ &0.008 & 0.013 & 0.021 & 0.443 & 1.448 & 4.175 \\
      &           & Stud & cp             &0.956 & 0.954 & 0.949 & 0.951 & 0.957 & 0.957 \\
      &           &      & $\overline{w}$ &0.008 & 0.013 & 0.021 & 0.443 & 1.448 & 4.175 \\
      &PAR        & VST  & cp             & 0.949 & 0.954 & 0.956 & 0.956 & 0.954 & 0.956 \\
      &           &      & $\overline{w}$ &0.002 & 0.005 & 0.014 & 1.534 & 8.372 & 39.466 \\
      &           & Stud & cp             & 0.949 & 0.954 & 0.956 & 0.955 & 0.955 & 0.956 \\
      &           &      & $\overline{w}$ &0.002 & 0.005 & 0.014 & 1.535 & 8.376 & 39.503 \\ \hline
10000   &LN         & VST  & cp             &0.956 & 0.952 & 0.954 & 0.953 & 0.953 & 0.958 \\
      &           &      & $\overline{w}$ &0.004 & 0.007 & 0.013 & 0.375 & 1.185 & 3.153 \\
      &           & Stud & cp             &0.956 & 0.953 & 0.954 & 0.953 & 0.953 & 0.957 \\
      &           &      & $\overline{w}$ &0.004 & 0.007 & 0.013 & 0.375 & 1.185 & 3.154 \\
      &$\chi^2_3$ & VST  & cp             &0.951 & 0.954 & 0.950 & 0.952 & 0.949 & 0.952 \\
      &           &      & $\overline{w}$ &0.006 & 0.009 & 0.015 & 0.312 & 1.018 & 2.925 \\
      &           & Stud & cp             &0.952 & 0.954 & 0.949 & 0.952 & 0.948 & 0.952 \\
      &           &      & $\overline{w}$ &0.006 & 0.009 & 0.015 & 0.312 & 1.018 & 2.925 \\
      &PAR        & VST  & cp             & 0.951 & 0.951 & 0.951 & 0.951 & 0.957 & 0.955 \\
      &           &      & $\overline{w}$ &0.002 & 0.004 & 0.010 & 1.079 & 5.890 & 27.727 \\
      &           & Stud & cp             & 0.950 & 0.951 & 0.951 & 0.951 & 0.957 & 0.954 \\
      &           &      & $\overline{w}$ &0.002 & 0.004 & 0.010 & 1.079 & 5.891 & 27.740 \\
       \end{tabular}
\end{footnotesize}
\end{table}

\begin{table}
\centering
\begin{footnotesize}
\caption{Coverage probabilities for the VST intervals using the lognormal QOR with  proportion of zeroes in the mixture distribution set to 0.01, 0.02 and 0.05. \label{table6}}
\vspace{0.5cm}
\begin{tabular}{rrrrrrrrr} $\epsilon$ & $n$ &   & 5/95 & 10/90 & 20/80 & 80/20 & 90/10 & 95/5 \\
  \hline
 0.01&  1000         &  LN          & 0.977 & 0.959 & 0.962 & 0.959 & 0.956 & 0.974 \\
     &               &  $\chi^2_3$  & 0.960 & 0.960 & 0.964 & 0.965 & 0.973 & 0.965 \\
     &               &  PAR         & 0.954 & 0.954 & 0.956 & 0.966 & 0.957 & 0.957 \\   \\
     &  5000         &  LN          & 0.967 & 0.958 & 0.950 & 0.962 & 0.961 & 0.968 \\
     &               &  $\chi^2_3$  & 0.967 & 0.944 & 0.958 & 0.952 & 0.951 & 0.951 \\
     &               &  PAR         & 0.949 & 0.957 & 0.956 & 0.960 & 0.948 & 0.961 \\  \\
     &  10000        &  LN          & 0.959 & 0.955 & 0.962 & 0.966 & 0.949 & 0.957 \\
     &               &  $\chi^2_3$  & 0.958 & 0.961 & 0.958 & 0.945 & 0.958 & 0.963 \\
     &               &  PAR         & 0.947 & 0.957 & 0.946 & 0.951 & 0.953 & 0.950 \\  \hline
 0.02&  1000         &  LN          & 0.974 & 0.976 & 0.960 & 0.967 & 0.968 & 0.979 \\
     &               &  $\chi^2_3$  & 0.950 & 0.968 & 0.959 & 0.966 & 0.961 & 0.956 \\
     &               &  PAR         & 0.939 & 0.950 & 0.954 & 0.963 & 0.962 & 0.934 \\  \\
     &  5000         &  LN          & 0.979 & 0.958 & 0.955 & 0.958 & 0.958 & 0.974 \\
     &               &  $\chi^2_3$  & 0.970 & 0.965 & 0.950 & 0.953 & 0.958 & 0.958 \\
     &               &  PAR         &  0.948 & 0.946 & 0.966 & 0.966 & 0.957 & 0.940 \\     \\
     &  10000        &  LN          & 0.976 & 0.958 & 0.953 & 0.945 & 0.957 & 0.974 \\
     &               &  $\chi^2_3$  & 0.952 & 0.947 & 0.954 & 0.950 & 0.952 & 0.965 \\
     &               &  PAR         &  0.963 & 0.952 & 0.956 & 0.950 & 0.950 & 0.938 \\ \hline
 0.05&  1000         &  LN          &       & 0.973 & 0.971 & 0.958 & 0.972 &       \\
     &               &  $\chi^2_3$  &       & 0.940 & 0.959 & 0.961 & 0.939 &       \\
     &               &  PAR         &       & 0.923 & 0.955 & 0.961 & 0.932 &       \\ \\
     &  5000         &  LN          &       & 0.982 & 0.954 & 0.953 & 0.973 &       \\
     &               &  $\chi^2_3$  &       & 0.967 & 0.960 & 0.945 & 0.972 &       \\
     &               &  PAR         &       & 0.947 & 0.936 & 0.955 & 0.930 &       \\  \\
     &  10000        &  LN          &       & 0.978 & 0.956 & 0.955 & 0.981 &       \\
     &               &  $\chi^2_3$  &       & 0.963 & 0.951 & 0.954 & 0.966 &       \\
     &               &  PAR         &       & 0.945 & 0.964 & 0.955 & 0.959 &       \\
\end{tabular}
\end{footnotesize}
\end{table}

\begin{table}
\begin{footnotesize}
\begin{center}
\caption{Australian equivalized weekly income (EWI) data for financial years beginning July 1, 2005 and July 1, 2011, in terms of 2011-2012 dollar values,  adjusted for the consumer price index. \cite{ABSincomedata}, Subset of Table on p.\;27, Document 6523.0, 2011-2012; downloaded 29/03/2015.\label{table7}}
 \vspace{0.5cm}
 \begin{tabular}{rrr}

    & \multicolumn{2}{c}{Number of persons ('000)}\\
   EWI  &   2005-2006 &   2011-2012 \\
\hline
 No income     &    73.7    &   87.4  \\
 \$1-\$49      &    90.1    &   83.7   \\
 \$50-\$99     &    66.7    &  101.8  \\
 \$100-\$149   &    76.3    &   88.2   \\
 \$150-\$199   &   121.9    &  121.5   \\
 \$200-\$249   &   259.0    &  225.9   \\
 \$250-\$299   &   710.3    &  382.3   \\
 \$300-\$349   &  1244.6    &  475.3    \\
 \$350-\$399   &  1235.7    & 1221.4    \\
 \$400-\$449   &  1139.8    & 1097.8    \\
 \$450-\$499   &  1070.7    & 1133.0    \\
 \$500-\$599   &  2189.4    & 2026.0    \\
 \$600-\$699   &  2259.2    & 2040.7    \\
 \$700-\$799   &  1922.5    & 2191.2    \\
 \$800-\$899   &  1647.9    & 1983.0    \\
 \$900-\$999   &  1350.6    & 1467.7    \\
 \$1000-\$1099 &  1048.9    & 1522.2    \\
 \$1100-\$1399 &  1847.3    & 2816.8    \\
 \$1400-\$1699 &   735.2    & 1484.1    \\
 \$1700-\$1999 &   334.7    &  713.4   \\
 \$2000 or more&   506.2    &  925.5   \\
\hline & {\bf 19930.7} & {\bf 22189.0} \\
        \end{tabular}
\end{center}
   \end{footnotesize}
   \end{table}

\begin{table}
\centering
\caption{Coverage probabilities (cp) and mean ($\overline{w}$) interval widths for the VST and Studentized intervals based on GLD QOR bandwidths.\label{table8}}
\vspace{0.5cm}
\begin{footnotesize}
\begin{tabular}{lrllrrrrrr} $n$ & $F$ &    &  & 5/95 &10/90 & 20/80 & 80/20 & 90/10 & 95/5 \\
  \hline
100   &LN         & VST  & cp             & 0.958 & 0.973 & 0.970 & 0.960 & 0.972 & 0.970 \\
      &           &      & $\overline{w}$ & 0.057 & 0.081 & 0.148 & 4.312 & 14.220 & 40.999 \\
      &           & Stud & cp             & 0.961 & 0.972 & 0.973 & 0.965 & 0.974 & 0.972 \\
      &           &      & $\overline{w}$ & 0.058 & 0.083 & 0.149 & 4.358 & 14.607 & 43.458 \\
      &$\chi^2_3$ & VST  & cp             & 0.954 & 0.959 & 0.954 & 0.960 & 0.965 & 0.958 \\
      &           &      & $\overline{w}$ & 0.063 & 0.096 & 0.157 & 3.533 & 12.161 & 39.830 \\
      &           & Stud & cp             & 0.963 & 0.967 & 0.969 & 0.956 & 0.969 & 0.954 \\
      &           &      & $\overline{w}$ & 0.068 & 0.099 & 0.159 & 3.541 & 12.241 & 40.366 \\
      &PAR        & VST  & cp             & 0.944 & 0.960 & 0.970 & 0.957 & 0.967 & 0.961 \\
      &           &      & $\overline{w}$ & 0.022 & 0.043 & 0.112 & 13.263 & 80.343 & 1048.191 \\
      &           & Stud & cp             & 0.952 & 0.966 & 0.967 & 0.963 & 0.970 & 0.963 \\
      &           &      & $\overline{w}$ & 0.025 & 0.045 & 0.114 & 13.449 & 83.968 & 1124.213 \\  \hline
250   &LN         & VST  & cp             & 0.968 & 0.966 & 0.967 & 0.967 & 0.965 & 0.969 \\
      &           &      & $\overline{w}$ & 0.032 & 0.049 & 0.089 & 2.607 & 8.358 & 22.975 \\
      &           & Stud & cp             & 0.965 & 0.964 & 0.965 & 0.967 & 0.966 & 0.970 \\
      &           &      & $\overline{w}$ & 0.032 & 0.050 & 0.089 & 2.616 & 8.427 & 23.358 \\
      &$\chi^2_3$ & VST  & cp             & 0.955 & 0.959 & 0.953 & 0.956 & 0.962 & 0.962 \\
      &           &      & $\overline{w}$ & 0.040 & 0.059 & 0.097 & 2.118 & 7.108 & 21.586 \\
      &           & Stud & cp             & 0.960 & 0.957 & 0.952 & 0.955 & 0.962 & 0.967 \\
      &           &      & $\overline{w}$ & 0.041 & 0.060 & 0.098 & 2.120 & 7.123 & 21.678 \\
      &PAR        & VST  & cp             & 0.951 & 0.954 & 0.951 & 0.960 & 0.959 & 0.959 \\
      &           &      & $\overline{w}$ & 0.011 & 0.025 & 0.066 & 7.511 & 42.221 & 223.830 \\
      &           & Stud & cp             & 0.957 & 0.953 & 0.950 & 0.968 & 0.961 & 0.968 \\
      &           &      & $\overline{w}$ & 0.011 & 0.026 & 0.066 & 7.543 & 42.732 & 230.261 \\ \hline
500   &LN         & VST  & cp             & 0.969 & 0.957 & 0.972 & 0.969 & 0.956 & 0.948 \\
      &           &      & $\overline{w}$ & 0.021 & 0.034 & 0.062 & 1.781 & 5.724 & 15.708 \\
      &           & Stud & cp             & 0.972 & 0.957 & 0.973 & 0.971 & 0.954 & 0.952 \\
      &           &      & $\overline{w}$ & 0.021 & 0.034 & 0.062 & 1.784 & 5.744 & 15.818 \\
      &$\chi^2_3$ & VST  & cp             & 0.961 & 0.950 & 0.958 & 0.954 & 0.969 & 0.957 \\
      &           &      & $\overline{w}$ & 0.028 & 0.042 & 0.068 & 1.467 & 4.775 & 14.209 \\
      &           & Stud & cp             & 0.958 & 0.950 & 0.957 & 0.957 & 0.970 & 0.957 \\
      &           &      & $\overline{w}$ & 0.028 & 0.042 & 0.069 & 1.467 & 4.780 & 14.235 \\
      &PAR        & VST  & cp             & 0.957 & 0.960 & 0.954 & 0.959 & 0.963 & 0.967 \\
      &           &      & $\overline{w}$ & 0.007 & 0.017 & 0.046 & 5.054 & 28.252 & 142.167 \\
      &           & Stud & cp             & 0.961 & 0.959 & 0.961 & 0.964 & 0.956 & 0.963 \\
      &           &      & $\overline{w}$ & 0.008 & 0.017 & 0.046 & 5.064 & 28.407 & 143.924 \\      \hline
       \end{tabular}
\end{footnotesize}
\end{table}

\begin{table}[h!]
\centering
\caption{Coverage probabilities for the interval estimators in \eqref{ci_diff} and, equivalently, \eqref{ci_diff2} comparing percentile ratios from the LN(0,1) and LN(0.2,1.5) distributions.  The sample sizes are $n$ and $m$ respectively.\label{table9}}
\begin{small}
\begin{tabular}{lllllll}\hline
$(n,m)$ & 5/95 & 10/90 & 20/80 & 80/20 & 90/10 & 95/5 \\ \hline
$(200,100)$ &0.976 &0.970    &0.970 &0.973 &0.972 &0.973   \\
$(500,1000)$ &0.969 &0.965   &0.963 &0.963 &0.964 &0.966   \\
$(10000,5000)$ &0.959 &0.957 &0.953 &0.952 &0.956 &0.960   \\ \hline
\end{tabular}
\end{small}
\end{table}

\clearpage
\newpage

\renewcommand{\baselinestretch}{1.5}

\begin{figure}
\begin{center}
\caption{Histograms of the data summarized in Table~\ref{table7}, after exclusion
of the first and last categories. Below each of them are density plots in solid lines of the reconstructed data sets described in the text. Superimposed in dashed lines are fitted gamma densities with  respective shape, scale parameters $(a,b)_{2005}=(3.94,184.88)$ and $(a,b)_{2011}=(4.23,197.61)$. \label{fig1}}
\end{center}
\end{figure}

\begin{figure}
\centering
\caption{Simulated coverage probability for the LOGN(0,1) distribution using the VST and studentized intervals for all combinations of $p$ and $q$ from $0.05, 0.06,\dots,0.95$.  1000 iterations were used for each combination.\label{fig2}}
\end{figure}

\begin{figure}
\centering
\caption{Simulated coverage probability for the Pareto(2) distribution using the VST and studentized intervals for all combinations of $p$ and $q$ from $0.05, 0.06,\dots,0.95$.  1000 iterations were used for each combination.\label{fig3}}
\end{figure}

\begin{figure}
\centering
\caption{Plots of $\IF (z;\,\rho _{p,q} ,F)/\rho _{p,q}(F)$ for which $z\in [0,1]$ and $p \in (0.05, 0.95)$ and $q=1-p$ (Plot A) and with $z=0$, $p \in (0.05, 0.95)$ and $q \in (0.05, 0.95)$ (Plot B).\label{fig4}}
\end{figure}

\begin{figure}
\centering
\caption{Plots of $\text{ASV}(\rho_{p,1-p};F)/\rho^2_{p,1-p}(F)$ for $p \in (0.05, 0.95)$ (Plot A) and with $p \in (0.05, 0.95)$ and $q = 1 - p$ (Plot B).\label{fig5}}
\end{figure}

\clearpage
\newpage

{\bf Figure 1}

\begin{figure}
\begin{center}
\includegraphics[scale=.7]{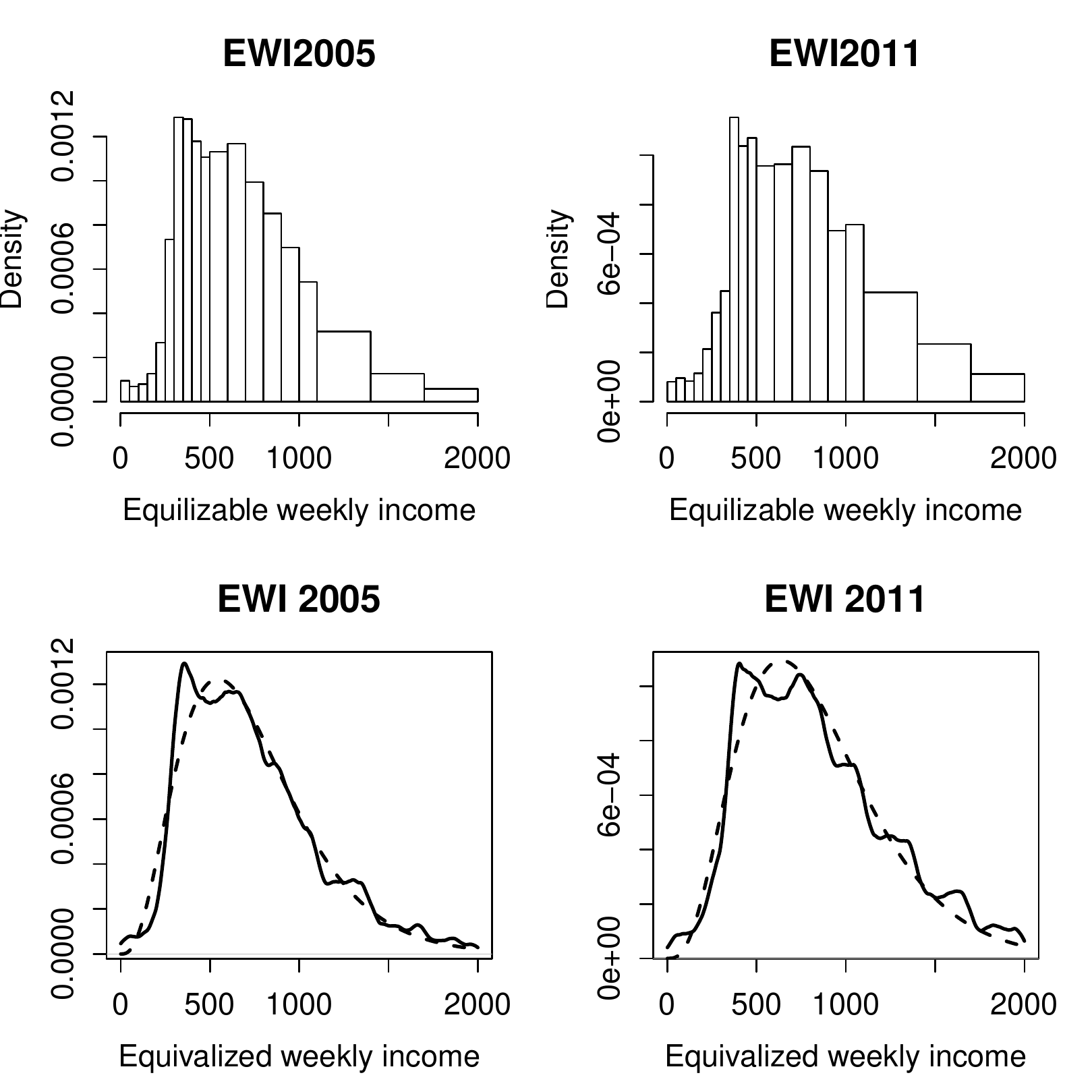}
\end{center}
\end{figure}

\clearpage
\newpage

\begin{figure}[t]
\centering
\includegraphics[scale=0.55]{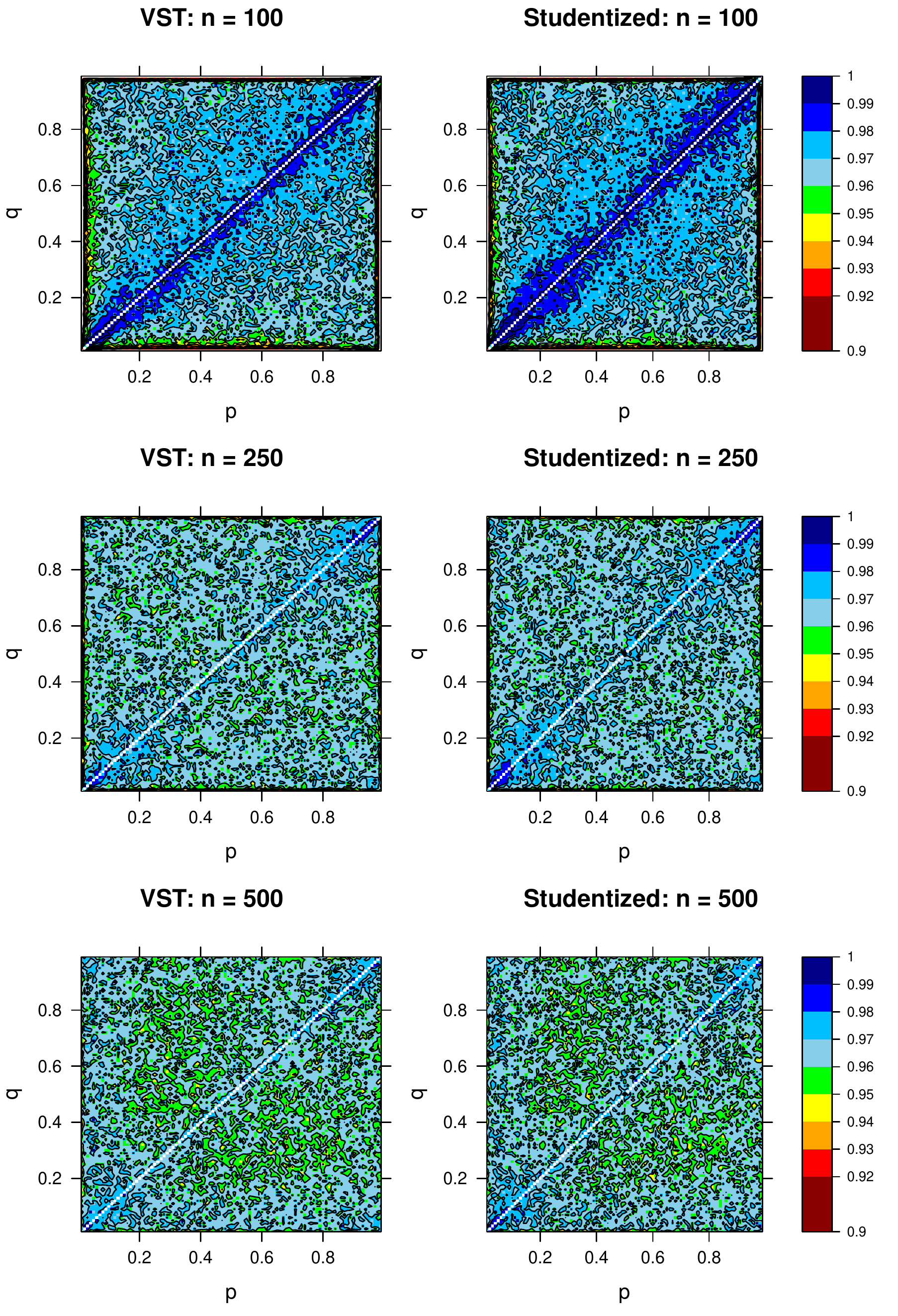}
\end{figure}
{\bf Figure 2}

\clearpage
\newpage

\begin{figure}[t]
\centering
\includegraphics[scale=0.55]{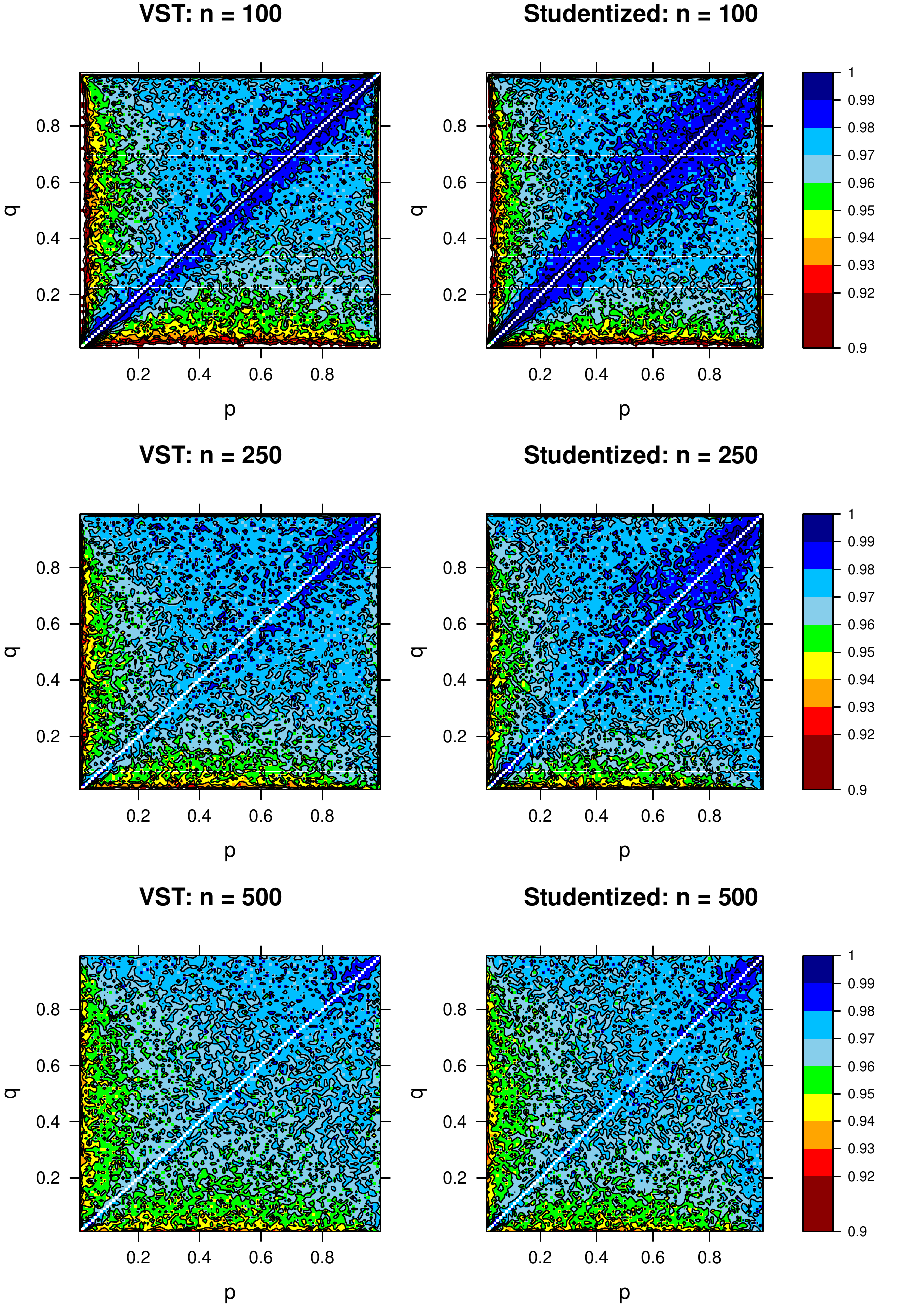}
\end{figure}
{\bf Figure 3}

\clearpage
\newpage

{\bf Figure 4}

\begin{figure}[t]
\centering
\includegraphics[scale=.6]{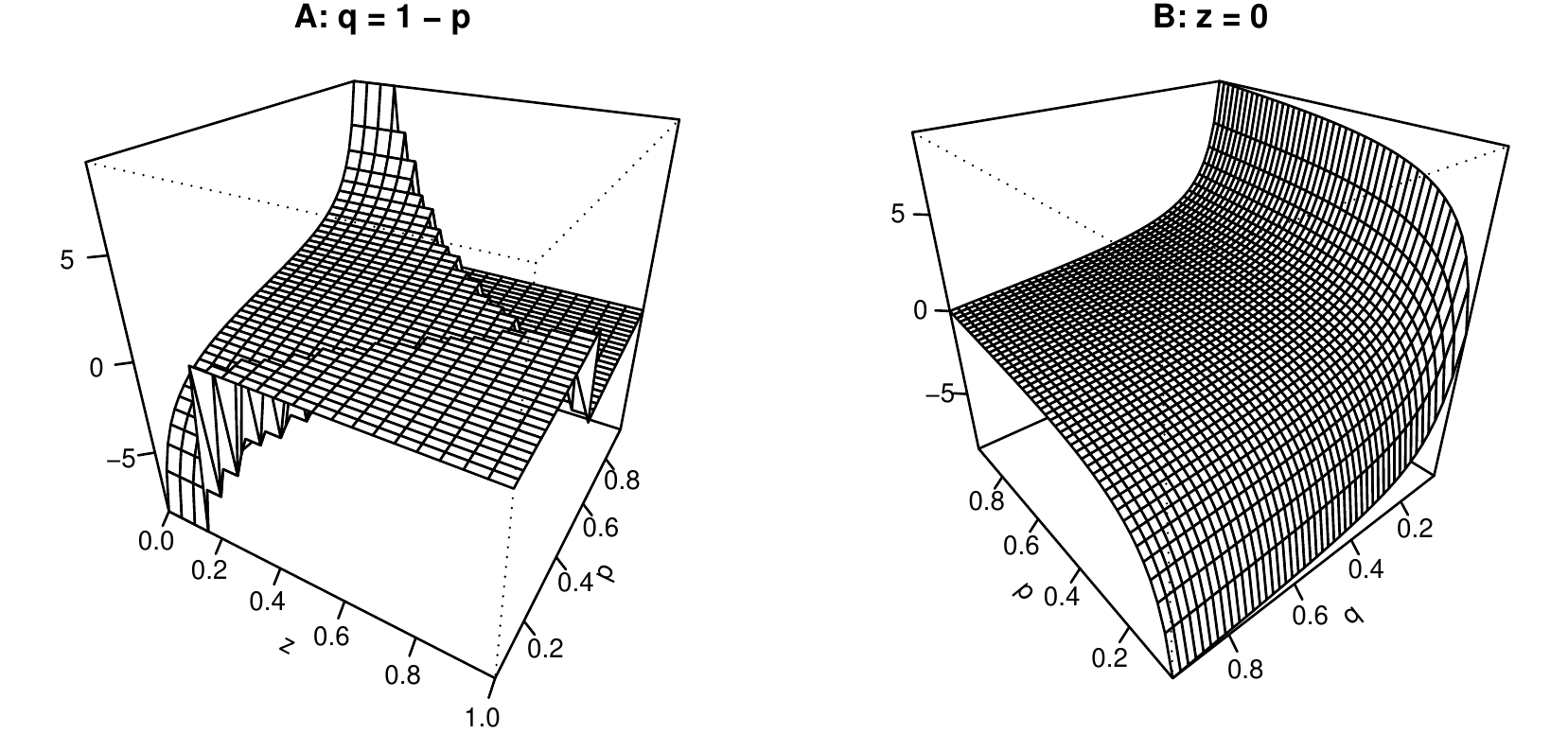}
\end{figure}

\vspace{2cm}

\begin{figure}[h]
\centering
\includegraphics[scale=.7]{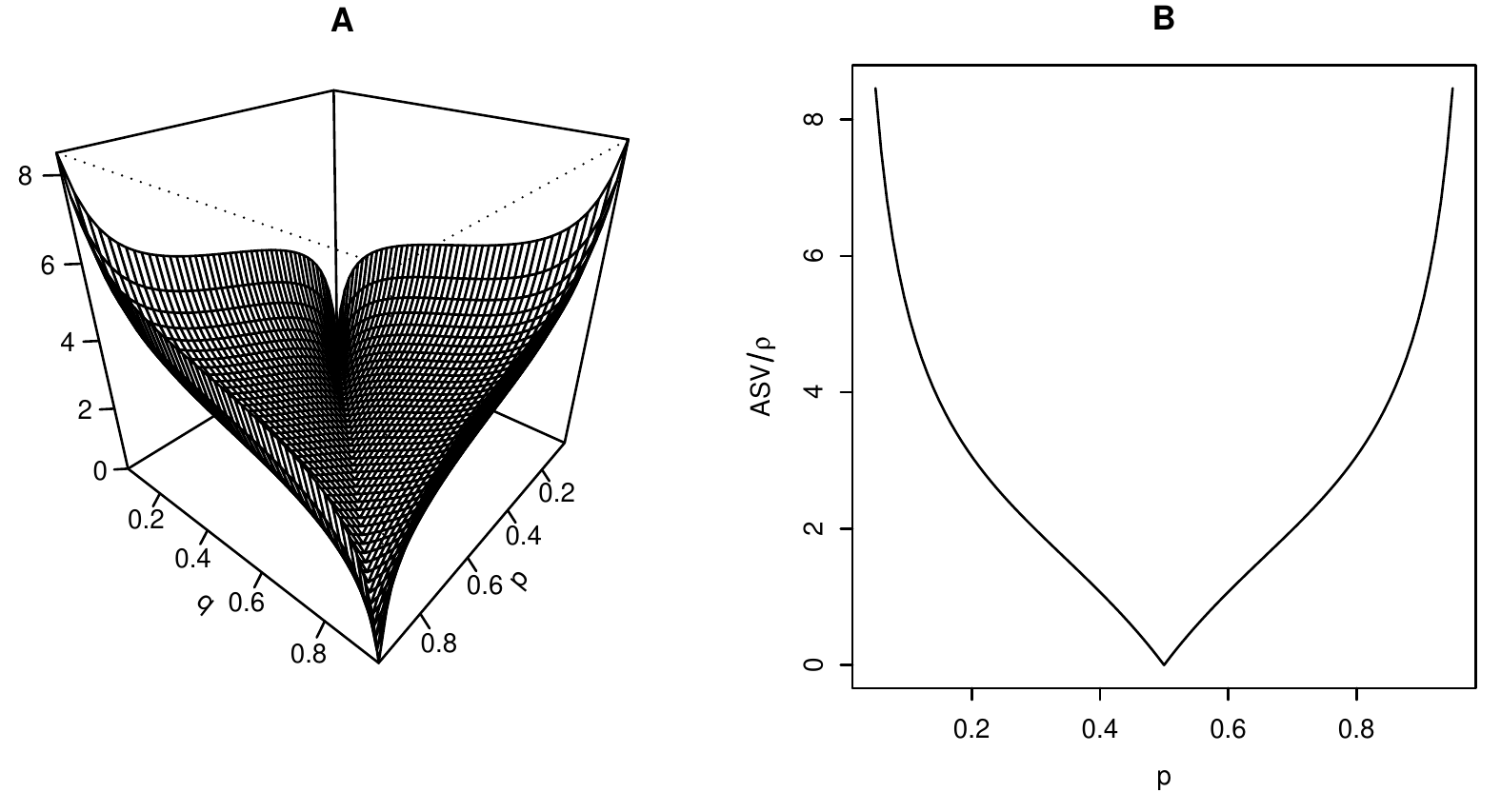}
\end{figure}

{\bf Figure 5}

\end{document}